\documentclass[aps,pra,epsfigure,twocolumn,superscriptaddress]{revtex4-2}
\usepackage[colorlinks=true,linkcolor=blue,urlcolor=blue,citecolor=blue,pdfusetitle]{hyperref}
\usepackage[english]{babel}
\usepackage{float}
\usepackage{afterpage}
\usepackage{amsmath}
\usepackage[caption = false]{subfig}
\usepackage{graphicx,epstopdf}
\usepackage{blindtext}
\usepackage[table,xcdraw,dvipsnames]{xcolor}
\usepackage{lipsum}
\usepackage{amsfonts}
\usepackage{bbm}
\usepackage{amssymb}
\usepackage{enumerate}
\usepackage{color}
\usepackage{latexsym}
\usepackage{physics}
\usepackage{times,txfonts}
\usepackage{soul}
\usepackage[normalem]{ulem}

\begin{document}

\title{Performance of quantum batteries with correlated and uncorrelated chargers}

\author{Mohammad B. Arjmandi}
\email{m.arjmandi@sci.ui.ac.ir}
\affiliation{Faculty of Physics, University of Isfahan, P.O. Box 81746-7344, Isfahan, Iran.}
\affiliation{Quantum Optics Research Group, University of Isfahan, Isfahan, Iran}

\author{Abbas Shokri} 
\affiliation{Physics Department, Azarbaijan Shahid Madani University, Tabriz, Iran}

\author{Esfandyar Faizi} 
\affiliation{Physics Department, Azarbaijan Shahid Madani University, Tabriz, Iran}

\author{Hamidreza Mohammadi}
\affiliation{Faculty of Physics, University of Isfahan, P.O. Box 81746-7344, Isfahan, Iran.}
\affiliation{Quantum Optics Research Group, University of Isfahan, Isfahan, Iran}

\date{\today}

\begin{abstract}
Energy can be stored in quantum batteries by electromagnetic fields as chargers. In this paper, the performance of a quantum battery with single and double chargers is studied. It is shown that by using two independent charging fields, prepared in coherent states, charging power of the quantum battery can be significantly improved, though the average number of embedded photons are kept the same in both scenarios. Then the results reveal that for the case of initially correlated states of the chargers the amount of extractable energy, measured by ergotropy, is more than initially uncorrelated ones, with appropriate degrees of field's intensities. Though the correlated chargers lead to greater reduction in purity of quantum battery, more energy and in turn, more ergotropy is stored in this case. In addition, we study the battery-charger mutual information and Von Neumann entropy and by using their relation, we find that both quantum and classical correlations are generated between the quantum battery and chargers. Then we study quantum consonance of the battery as the non-local coherence among it's cells and find some qualitative relations between the generation of such correlations and the capability of energy storage in the quantum battery.
\end{abstract}

\maketitle

\section{Introduction}
Batteries as the portable energy storage devices play a significant role in our modern life. Nearly all aspects of this new
era from household appliance to medical instruments, navigation, transportation and so on depend, in a way, on the performance of these systems. A typical battery contains one or more electrochemical cells in which an energy conversion occurs from chemical type to
electrical one, by means of the so-called reduction-oxidation reactions ~\cite{schmidt2018batteries,weber2011redox,pan2015redox}.\\
On the other hand, quantum effects become important as the size scale of electronic devices is quickly decreasing. So there may be an advantage from such quantum effects upon which one may build devices that show a better performance than their classical analogous ~\cite{geppert2000quantum,joulain2016quantum,mcrae2019graphene,shukla2008nonlinear,shen2014quantum,lvovsky2009optical,hedges2010efficient,dennis2002topological,ladd2010quantum,valiev2005quantum,divincenzo1999quantum}. One of such devices can be quantum storage system or quantum battery (QB).\\
QB is a collection of one or more quantum system(s) (mostly a two-level one) with the ability of energy storage. The notion of QB has been brought into the spotlight zone after the work by Alicki and Fannes ~\cite{Alicki:13}.\\
Generally, the study of QB performance has been divided into two distinct scenarios: charging and (self) discharging. The latter indicates a situation in which QB loses its energy due to interaction with surrounding environment ~\cite{arjmandi2022enhancing,Kamin:20-1,Santos:21b}. Actually, this is also prevalent among the traditional batteries ~\cite{conway1997diagnostic,Ricketts:00}. In the former, usually an external charging field acts as a charger in order to store energy in QB ~\cite{PRB2019Batteries,PRL_Andolina}.\\
Some figures of merit are proposed by which one can evaluate the useful capabilities and performance of a special QB. The most studied quantities are stored energy ~\cite{chang2021optimal,PRE2019Batteries}, ergotropy ~\cite{tabesh2020environment,Le:18,PRL_Andolina}, charging power ~\cite{Crescente:20,Andolina:19,ghosh2021fast} and so on.\\
Firstly, Alicki and Fannes claimed that global entangling operations end in an increment of energy extractable from QB ~\cite{Alicki:13}. However, this idea appeared to be imperfect since Hovhannisyan et al. ~\cite{PRL2013Huber} proved that the optimal energy extraction can be achieved using indirect consecutive permutation operations by which no entanglement is dynamically generated. But it is worth noting that the performance of such operations is a time consuming process. In fact, it is argued that collective charging operations (with entanglement creation) might offer a considerable speed up to charging process compared to parallel charging scenario in which each cell of QB is individually charged ~\cite{Binder:15}.\\
Charging power can also be enhanced by disordered interactions between quantum cells of QB ~\cite{ghosh2020enhancement}. Moreover, the effect of disorder and localization on QBs is studied ~\cite{Andolina:19-2}. It is shown that the batteries which are in many-body localized phase gain more stability and shorter optimal time scale of charging process, in comparison with those being the ergodic and Anderson localized phases.\\
 The role of quantum entanglement and coherence for two and three-cell QB are investigated in ~\cite{Kamin:20-2}. The authors have shown that while entanglement seems to be effectless or even destructive, quantum coherence presents a qualitative relation with the efficiency of the QB. However, more efforts are required to address the impact of quantum correlations in this field since a universal relation between the performance of quantum battery and its correlation contents has not yet been found ~\cite{araya2019geometrical}.\\
As stated before, energy storage in QB can be done using an external charging field. Andolina et al. considered three well-known quantum optical states Fock, coherent and squeezed vacuum as the initial state of the charger ~\cite{PRL_Andolina}. They confirmed that the coherent state is optimal in the sense of energy deposition and extraction because it results in a smaller amount of charger-battery entanglement.\\
In this paper, we consider a QB including four two-level quantum cells (qubit) interacting with one or two photonic cavities as charger(s), each of which prepared in a coherent initial state. The results reveal that, regarding the charging power of the QB, employing two independent chargers is significantly superior to single charger case, while the average photon number is kept the same for both cases. Furthermore, quantum correlated bipartite chargers eg. semi bell states result in more amount of ergotropy in comparison with uncorrelated one. Moreover, we study the purity of QB and find out that the QB with initial correlated chargers becomes more mixed than that with uncorrelated ones during charging process. This more mixedness in turn results in more energy storage in the QB. In addition, we numerically prove that the QB and its charging fields can share both quantum and classical correlations with the same values. Also the results show that such correlations between the QB and chargers along with the quantum consonance of the QB have some role to play as they positively interplay with the performance of QB.\\
This paper is organized as follows. The next section presents the physical model and its dynamics. Section III is dedicated to definition of some figures of merit and concepts which are used in this paper. In section IV the results of different charging scenarios and some discussions can be found. Finally, section V concludes the paper.
\section{Model}
In this paper, we consider an array of identical two-level quantum systems (qubit) as a QB, interacting with one or two independent single-mode photonic cavities as the charging device ~\cite{dicke1954coherence,tavis1968exact,tavis1969approximate}:

\begin{align}
H=H_{B}+H_{C}+H_{I},\label{eq1}
\end{align}
where
\begin{subequations}\label{eq:litdiff}
\begin{align}
H_{B} &=\frac{\omega_{0}}{2}\sum_{i}^{N_{B}}{\sigma_{i}^{z}} \label{eq2(a)}\\
H_{C} &=\sum_{j}^{N_{C}}{\omega_{j} a_{j}^{\dagger} a_{j}} \label{eq2(b)}\\
H_{I} &=\sum_{i,j}^{ N_{B},N_{C}}{ g_{ij}(\sigma_{i}^{+} a_{j}+\sigma_{i}^{-} a_{j}^{\dagger})} \label{eq2(c)}
\end{align}
\end{subequations}
are battery, charger and interaction Hamiltonians, respectively. In these equations, $N_{B}$ and $N_{C}$ are the number of quantum cells of the QB and the chargers respectively, $\sigma_{i}^{k} (k=x,y,z)$ denotes the well-known Pauli matrices and $\sigma_{i}^{\pm}$ are raising and lowering operators for ith qubit and $a_{j}^{\dagger}$ ($a_{j}$) are creation and annihilation operators of jth charger. $\omega_{0}$ is the energy splitting between the ground and excited states of each qubit while $\omega_{j}$ is the frequency of photons in each cavity and  $g_{ij}$ is the coupling strengths between ith qubit and jth charger.  Here, we consider the resonant condition $\omega_{j}=\omega_{0}$ and also assume that the coupling strengths are independent of battery and chargers i.e. $g_{ij}=g=2\omega_{0}$ which in turn means that we focus on strong coupling regimes ~\cite{yoshihara2017superconducting,langford2017experimentally,braumuller2017analog}. Suppose the QB and chargers undergo a unitary evolution, $U=exp(-iHt)$ where $H$ is given by Eq. 1 and for simplicity we set $\hbar = 1$:
\begin{align}
\rho_{B/C}(t)=Tr_{C/B}(U \rho_{B}(0)\otimes\rho_{C}(0) U^{\dagger} ),\label{eq3}
\end{align}
where $\rho_{B}(0)$ and $\rho_{C}(0)$ are initial states of battery and charger(s) respectively. Thorough this paper, we focus on an initial 4-cell empty battery in which all qubits are prepared in their ground states so $\rho_{B}(0)= (\ket{g} \bra{g})^{\otimes 4}$. Meanwhile, we set each charger to be in a coherent optical state $\ket{\alpha}$ where $\alpha$ is a complex number identifying the average number of embedded photons in the cavity as $\overline{n}=|\alpha|^2$ ~\cite{Scully:Book}. In fact, it is shown that coherent state is the optimal state to charge a quantum battery since it causes less amount of battery-charger entanglement ~\cite{PRL_Andolina}.
\section{Figures of merit}\label{sec3}
In this section, we provide some figures of merit by which one can evaluate the performance of a QB.\\
\subsection{Thermodynamical performance}
 Assume a quantum battery with density matrix $\rho$ and Hamiltonian $H$. Then its total amount of stored energy is simply given by:
\begin{align}
E(\rho,H)=Tr\:(\rho H).\label{eq3}
\end{align}
Mostly, due to different reasons, one is not able to entirely extract this energy through unitary operations. So we need to consider ergotropy i.e. the maximum amount of energy which can be unitarily extracted ~\cite{Allahverdyan:04}. It is defined as the difference between internal energies of quantum state $\rho$ and its corresponding passive state $\eta$:
\begin{align}
\mathcal E(\rho,H)=E(\rho,H) - E(\eta,H).\label{eq5}
\end{align}

\begin{figure}[t!]
\includegraphics[width=\linewidth]{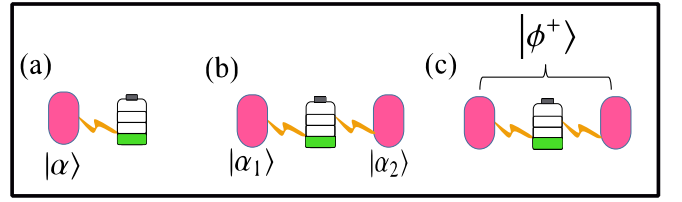}
\caption{Schematic figure of the three charging schemes: (a) single, (b) double uncorrelated and (c) double correlated chargers.}
\label{schematic}
\end{figure}

If $\eta$ is passive, it provides no energy by means of unitary operations. Passive state of a quantum system possesses non-increasing population with respect to its Hamiltonian $H$ and also $[H,\eta]=0$. Consider the spectral decompositions of the battery state and Hamiltonian as $\rho = \sum_{i=1}^{d} {p_{i} \ket{p_{i}}\bra{p_{i}}}$ and $H = \sum_{i=1}^{d} {\epsilon_{i} \ket{\epsilon_{i}}\bra{\epsilon_{i}}}$ so that $p_{1}\geq p_{2}\geq . . . \geq p_{d}$ and $\epsilon_{1}\leq \epsilon_{2}\leq . . . \leq \epsilon_{d}$ where $d$ denotes the dimension of the Hilbert space. Then the passive state is $\eta=\sum_{i=1}^{d} {p_{i} \ket{\epsilon_{i}}\bra{\epsilon_{i}}}$ and ergotropy can be written as
\begin{align}
\mathcal E(\rho,H)=\sum_{i,j}^{d} {p_{i} \epsilon_{j} (|\langle p_{i}|\epsilon_{j}\rangle|^{2} - \delta_{i,j})}.\label{eq6}
\end{align}

\begin{figure}[t!]
	\centering
		\includegraphics[width=\linewidth]{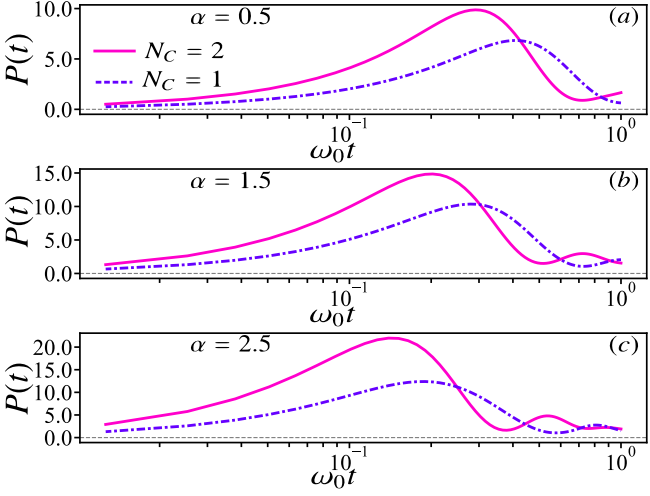}
	\caption{Charging power of the QB with one ($N_{C}=1$) and two ($N_{C}=2$) independent chargers, respectively in states $\ket{\alpha}$ and $\ket{\alpha_{1}}\ket{\alpha_{2}}$ with condition $|\alpha|^{2}=|\alpha_{1}|^{2}+|\alpha_{2}|^{2}$ with various amount of field's intensity.}
	\label{fig:pow}
\end{figure}

where $\delta_{i,j}$ is the Kronecker delta function. As ergotropy is the maximum amount of extractable energy using unitary operations so the most successful extracting operations are those which transform the quantum state $\rho$ to the passive state $\eta$.\\
Another important figure of merit that determines how fast a quantum battery can be charged is power:
\begin{align}
P(t)=\frac{E(\rho_{B}(t),H)}{t},\label{eq8}
\end{align}
where $t$ denotes the charging time.
\subsection{Correlations}
As previously mentioned, the most efforts are dedicated to the role of quantum entanglement in the performance of quantum batteries. It appears to be effective \cite{James:20,tabesh2020environment}, destructive \cite{PRL_Andolina,tabesh2020environment} or effectless \cite{Kamin:20-2,ghosh2021fast}. So it makes sense to seek another quantifier of quantum correlations which may show more correspondence with performance of quantum batteries. Moreover, entanglement is not the only quantifier of quantum correlations. In this paper, we consider quantum consonance which estimates the global coherence in quantum systems and for a general multipartite density matrix is defined by ~\cite{pei2012using}
\begin{align}
C(\rho)=\sum_{k_{1}k_{2}...k_{n},l_{1}l_{2}...l_{n}}{|\rho_{k_{1}k_{2}...k_{n},l_{1}l_{2}...l_{n}}^{c}\prod_{m}{(1-\delta_{k_{m},l_{m}})}|},
\end{align}
where $\rho^{c}=U \rho U^{\dagger}$ is the transformed density matrix by using some unitary operations which eliminate the local coherence of $\rho$. So unlike entanglement, quantum consonance can be easily calculated even for multipartite quantum systems. For a two-qubit system $\rho_{AB}$ the above relations reduces to
\begin{align}
C(\rho_{AB})=\sum_{k_{1}k_{2},l_{1}l_{2}}{|\rho_{k_{1}k_{2},l_{1}l_{2}}^{c}(1-\delta_{k_{1},l_{1}}) (1-\delta_{k_{2},l_{2}})|},
\end{align}

\begin{figure}[t!]
	\centering
		\includegraphics[width=\linewidth]{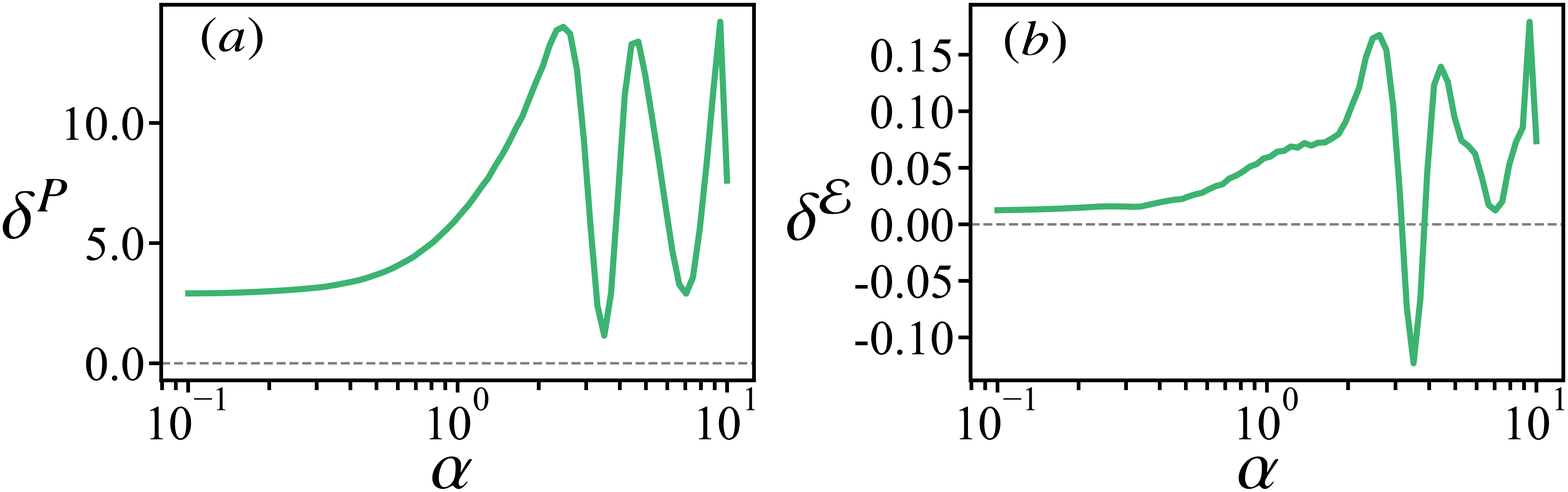}
	\caption{Behavior of (a) $\delta^{\mathcal E}$ and (b) $\delta^{P}$ versus the degree of field's intensity.}
	\label{fig:dif}
\end{figure}

where $\rho_{k_{1}k_{2},l_{1}l_{2}}^{c}=(U_{A} \otimes U_{B}) \rho_{AB} (U_{A} \otimes U_{B})^{\dagger}$ in which $U_{A}$ and $U_{B}$ are some transformations that remove the local coherence of the reduced states $\rho_{A/B}=Tr_{B/A}\: (\rho_{AB})$. The subindices $k_{1},l_{1}$ and $k_{2},l_{2}$ label the basis of the two qubits as $\{ \ket{00}, \ket{01}, \ket{10}, \ket{11} \}$. Using Eq. 9, it can be easily shown that quantum consonance includes only the elements related to $\ket{00} \bra{11}$, $\ket{01} \bra{10}$, $\ket{10} \bra{01}$ and $\ket{11} \bra{00}$ of the transformed state $\rho_{k_{1}k_{2},l_{1}l_{2}}^{c}$. The generalization to the four-qubit case is also straightforward.
  \\
Moreover, one may want to know how much the battery and charger(s) become correlated during the charging process. It can be quantified by battery-charger mutual information. It evaluates the total amount of correlations (classical+quantum) between two parts. Let us consider $\rho_{BC}$ to be the joint density matrix of battery and charger. Then mutual information is given by ~\cite{mahdian2016comparison,Ollivier:01}
 
\begin{align}
I(\rho_{BC})=\chi(\rho_{BC})+Q(\rho_{BC})=S(\rho_{B})+S(\rho_{C})-S(\rho_{BC}),\label{mutual}
\end{align}

in which $\chi(\rho_{BC})$ and $Q(\rho_{BC})$ denote the classical and quantum correlation between the QB and chargers, respectively and $S(\rho)=Tr\:(\rho\: log_{2}\: \rho)$ is the Von Neumann entropy. 
\section{Two scenarios: charging with single or double photonic cavities }\label{sec4}
As we previously mentioned, we consider each charger to be in coherent state $\ket{\alpha}$. In this section, we investigate two distinct charging scenarios (Fig.~\ref{schematic}): charging with $i)$ one single-mode cavity $\ket{\alpha}$ and $ii)$ two single-mode cavities $\{\ket{\alpha_{1}},\ket{\alpha_{2}}\}$ with different configurations (correlated or uncorrelated).

\begin{figure}[h!]
	\centering
		\includegraphics[width=\linewidth]{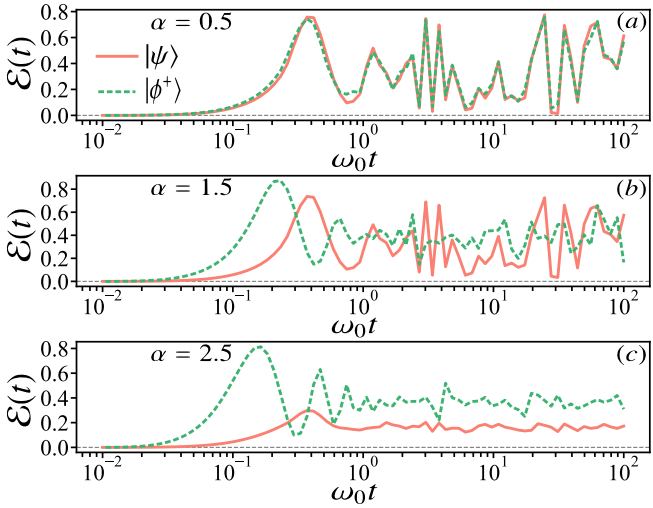}
	\caption{Ergotropy of the QB with initially correlated ($\ket{\phi^{+}}$) and uncorrelated ($\ket{\psi}$) chargers for (a) $\alpha=0.5$, (b) $\alpha=1.5$ and (c) $\alpha=2.5$.}
	\label{fig:erg}
\end{figure}

\subsection{Boosting the charging power}
Now, we compare two charging cases in the first of which there is only one charger ($N_{C}=1$) in state $\ket{\alpha}$, while the second case contains two uncorrelated chargers ($N_{C}=2$) in state $\ket{\alpha_{1}}\ket{\alpha_{2}}$. For an unbiased comparison between these charging schemes we consider an equal average number of photons in both cases ($|\alpha|^{2}=|\alpha_{1}|^{2}+|\alpha_{2}|^{2}$). Fig.~\ref{fig:pow} shows the charging power for these two charging protocols with $ \alpha=0.5, 1.5,$ and $ 2.5$. Clearly, in all cases, charging with two chargers appears to have a better performance in terms of power, regardless of field intensities. It is an interesting result: though the average number of photons in both charging approaches is the same, it is possible to charge the battery with a higher rate if one use two cavities. Moreover, raising the average number of embedded photons in the cavities leads to the growth of maximum power. In fact, the gain in ability of faster charging in double cavities compared to the single cavity can be up to $76$ percent for $\alpha=2.5$. One may think of a monotonic enhancement of charging power for double chargers with respect to the single one when $\alpha$ is arbitrarily increased. However, this is not the case because we plot the maximum of charging power versus the field's intensity in Appendix ~\ref{appendix:a} and it shows a periodic dependence on $\alpha$. \\
 In addition, the double charger case needs a shorter time scale to reach the maximum power compared to the single charger scheme for which the peaks take longer times to emerge.\\ 
We should mention that single and double uncorrelated charger show no considerable advantage over each other, regarding the ergotropy storage though the latter being slightly better. To show this we consider the difference between the maximum amount of ergotropy over time for these charging schemes i.e. $\delta^{\mathcal E}=\mathcal E _{max}^{N_{C}=2}-\mathcal E _{max}^{N_{C}=1}$ where $\mathcal E _{max}^{N_{C}=2}$ and $\mathcal E _{max}^{N_{C}=1}$ are the maximal ergotropy of QB with double uncorrelated and single chargers respectively. We define the same quantity for charging power as $\delta^{P}$. In Fig. ~\ref{fig:dif} we plot these differences versus the degree of field's intensity. While $\delta^{P}$ is considerably large for various amounts of $\alpha$ (Fig. ~\ref{fig:dif}a), as expected from Fig. ~\ref{fig:pow}, $\delta^{\mathcal E}$ is small and even negative for some field's intensities (Fig. ~\ref{fig:dif}b) denoting $\mathcal E _{max}^{N_{C}=1} > \mathcal E _{max}^{N_{C}=2}$.\\
In what follows, we want to see if different configurations of chargers with or without correlations play any significant role in the charging of the QB.
\subsection{Correlated and uncorrelated chargers}
One may ask if the existence of correlations between chargers can result in a considerable effect during the charging process. So here we aim at addressing this point. We consider two double chargers schemes. In the first one, the state of two chargers is a product state $\ket{\psi}=\ket{\alpha_{1}}\ket{\alpha_{2}}$, while in the second, the two chargers are prepared to be in an entangles semi Bell state ~\cite{wang2001multipartite,chatterjee2021quantifying}

\begin{figure}[t!]
	\centering
		\includegraphics[width=\linewidth]{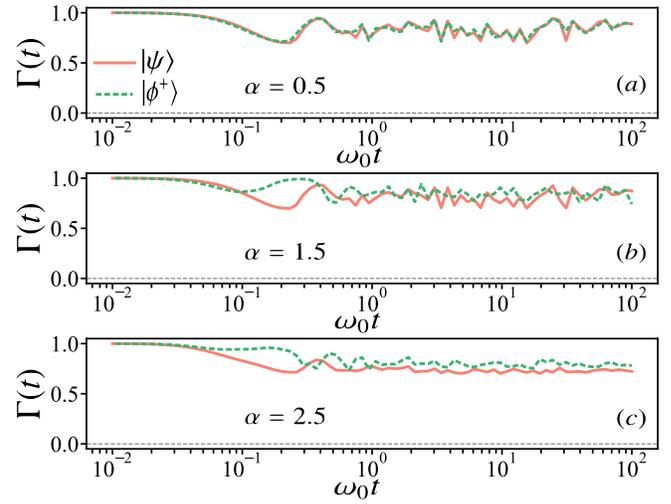}
	\caption{Ratio of extractable energy of the QB with initially correlated and uncorrelated chargers with the same parameters as Fig.~\ref{fig:erg}.}
	\label{fig:eff}
\end{figure}

\begin{equation}
\ket{\phi^{+}}=\frac{1}{\sqrt{N_{+}}}(\ket{\alpha_{1}}\ket{\alpha_{1}}+\ket{\alpha_{2}}\ket{\alpha_{2}}),\label{eq8}
\end{equation}
where $N_{+}=2(1+e^{-2(|\alpha_{1}|^{2}+|\alpha_{2}|^{2})})$ is normalization constant. The appellation of such states as semi Bell, quasi Bell or Bell-like states comes from their similarity to the well-known Bell states $\ket{\phi^{\pm}}=(\ket{00} \pm \ket{11})/\sqrt{2}$ and $\ket{\psi^{\pm}}=(\ket{01} \pm \ket{10})/\sqrt{2}$ in which $\{ \ket{00},\ket{01},\ket{10},\ket{11} \}$ are the standard orthogonal basis of a two-qubit system. However, the semi Bell states are made of continuous variable states, namely coherent states without orthogonality condition i.e. $\bra{\alpha_{1}}\ket{\alpha_{2}} \neq 0$ for $\alpha_{1} \neq \alpha_{2}$. Various schemes of their experimental preparation have been proposed \cite{solano2003strong,paternostro2003generation,jeong2006greenberger}. Through the rest of paper, we set $\alpha_{1}=-\alpha_{2}=\alpha$. In Appendix.~\ref{appendix:c} we provide the results of other choices of semi Bell states to prove the universality of our results.

\begin{figure}[h!]
	\centering
		\includegraphics[width=\linewidth]{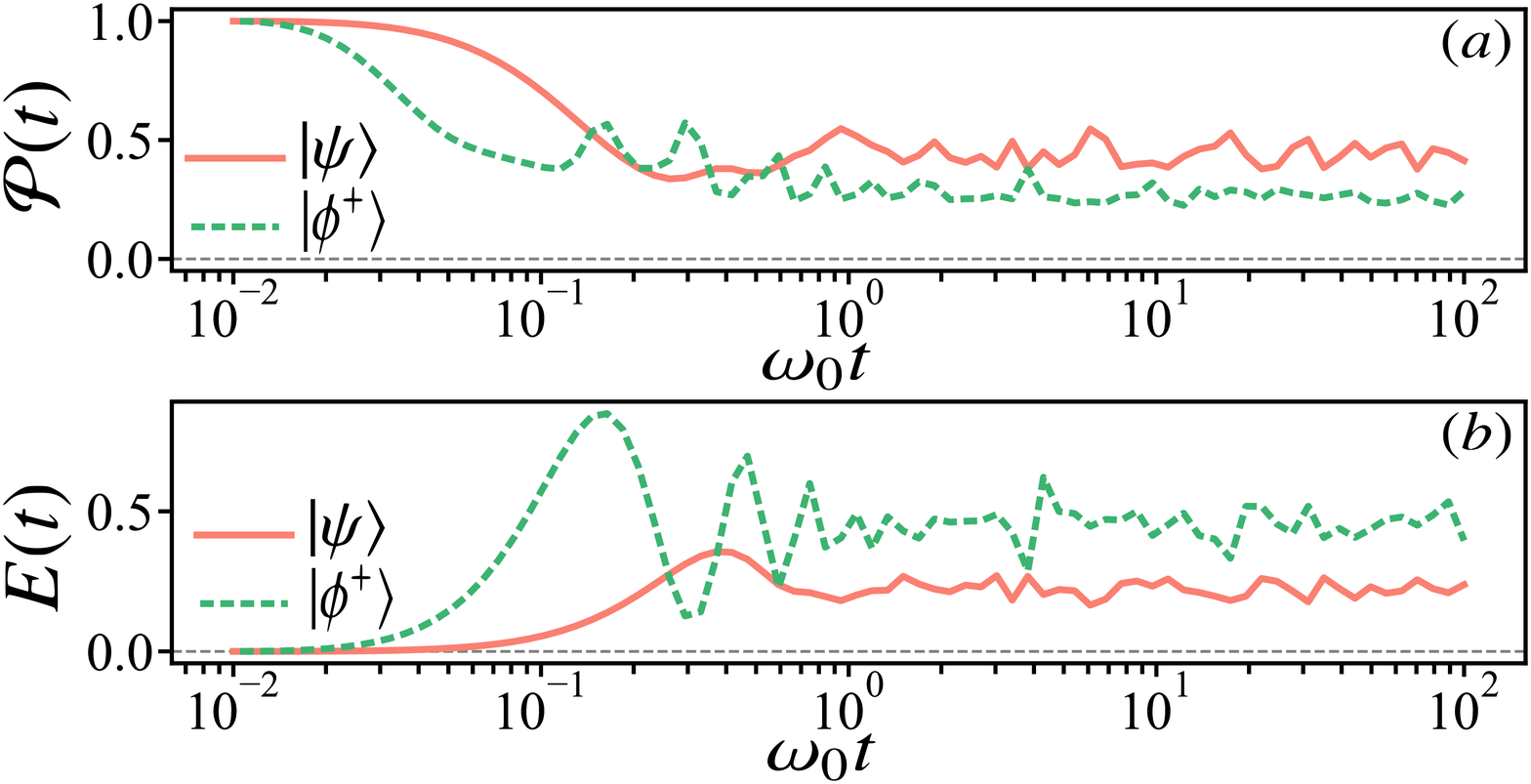}
	\caption{ The dynamics of (a) purity and (b) stored energy of QB with the two states $\ket{\psi}$ and $\ket{\phi^{+}}$ as the chargers for $\alpha=2.5$ .}
	\label{fig:purity-energy}
\end{figure}

\begin{figure}[h!]
	\centering
		\includegraphics[width=\linewidth]{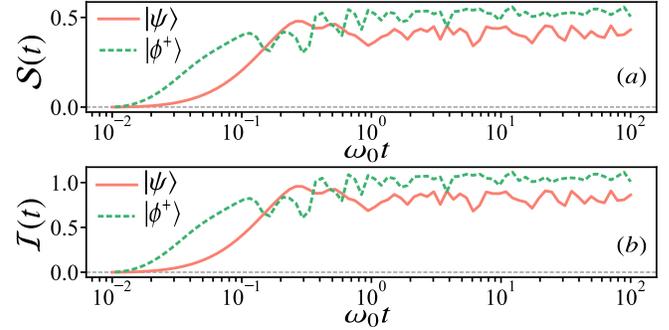}
	\caption{ Dynamics of battery-charger (a) entanglement and (b) mutual information for $\ket{\psi}$ and $\ket{\phi^{+}}$ with $\alpha=2.5$.}
	\label{fig:ent-von}
\end{figure}

Fig.~\ref{fig:erg} shows the ergotropy of quantum battery i.e. $\mathcal E (t)=\mathcal E (\rho_{B}(t),H_{B})/N_{B}\omega_{0}$ versus time for the two initial states $\ket{\psi}$ and $\ket{\phi^{+}}$ of the cavities. A significant capability of the chargers in semi Bell state in charging the quantum battery can be inferred when $\alpha$ is large enough, compared to the uncorrelated state $\ket{\psi}$ (Fig.~\ref{fig:erg}c). In fact, when the cavities share no initial correlation, raising the intensities of the charging field reduces the overall stored ergotropy in the quantum battery, though there is no considerable difference between the two types of chargers for smaller values of $\alpha$ (Fig.~\ref{fig:erg}a). However, the dependence of maximum ergotropy on the degree of field's intensity ($\alpha$) is periodic as shown in Appendix.~\ref{appendix:a} . In addition, it can be qualitatively seen that the maximum amount of ergotropy takes a shorter time to be deposited in the quantum battery (Fig.~\ref{fig:erg}b and \ref{fig:erg}c) if one uses an initially correlated state for the photonic cavities. Therefore, this kind of charging scheme is also able to provide some advantages in terms of charging power.\\
The ratio of ergotropy and internal energy i.e. $\Gamma (t)=\frac{\mathcal E(\rho_{B} (t),H)}{E(\rho_{B} (t),H)}$  is also an interesting quantity since it determines what fraction of the total stored energy is extractable. In Fig.~\ref{fig:eff} this ratio is depicted versus time for the two types cavity initial states. A similar behavior to ergotropy can be seen from this figure. While the ratios obtained by the two charging settings are almost the same for $\alpha=0.5$, increasing the intensity of charging fields results in the distinction of the extractable energy ratio deposited in the quantum battery by the correlated cavities.\\
Now let us take a careful look at the purity of QB defined by
\begin{equation}
\mathcal P (t)=Tr\: [\rho_{B}(t)^{2}].\label{purit}
\end{equation}
For the sake of brevity, we only consider the result with $\alpha=2.5$ for which more distinction between the two considered charging schemes emerges. As shown by Fig.~\ref{fig:purity-energy}, the purity of QB turns out with a somehow opposite behavior to ergotropy (Fig. ~\ref{fig:erg}c). Here, the separable chargers keep the QB to possess larger purity during the time evolution, with respect to the correlated ones while this is not the case for ergotropy. On the other hand, it is known that if a quantum state is pure, it can be unitarily transformed into its ground state. So, by using Eq. 5 we have $\mathcal E (\rho_{B},H_{B})=E(\rho_{B},H_{B})$ i.e. the whole of energy can be extracted in form of ergotropy (with the assumption of zero energy for the ground state). But for mixed states $\mathcal E (\rho_{B},H_{B}) < E(\rho_{B},H_{B})$ \cite{PRL_Andolina}. Consequently, QBs with larger purity also offer larger extractable energy. One may think this is violated by our results but this is not true. In fact, for a given energy, the more purity QB contains, the closer its ergotropy will be to its energy. In our cases, different charging schemes result in different amounts of stored energy as seen from Fig.~\ref{fig:purity-energy}b. Obviously, the stored energy in the QB with entangled chargers is more than that for the separable ones and this leads, in turn, to the greater amount of ergotropy.\\
In addition, the higher purity of the QB with separable chargers is due to less generated battery-charger entanglement. Fig. ~\ref{fig:ent-von}a shows the Von Neumann entropy of the QB i.e. $\mathcal S (t)=S(\rho_{B}(t))/N_{B}$ as the measure of total system (QB+chargers) entanglement for $\alpha=2.5$. Clearly, the QB becomes more entangled (and mixed, in turn)  with correlated chargers than separable ones. Overall, the battery-charger entanglement appears with two distinct roles: it results in more energy injection into the QB and consequently more ergotropy but it allows smaller fractions of energy to be extractable. In our work, at least, the former role tends to be dominant.\\

\begin{figure}[h!]
	\centering
		\includegraphics[width=\linewidth]{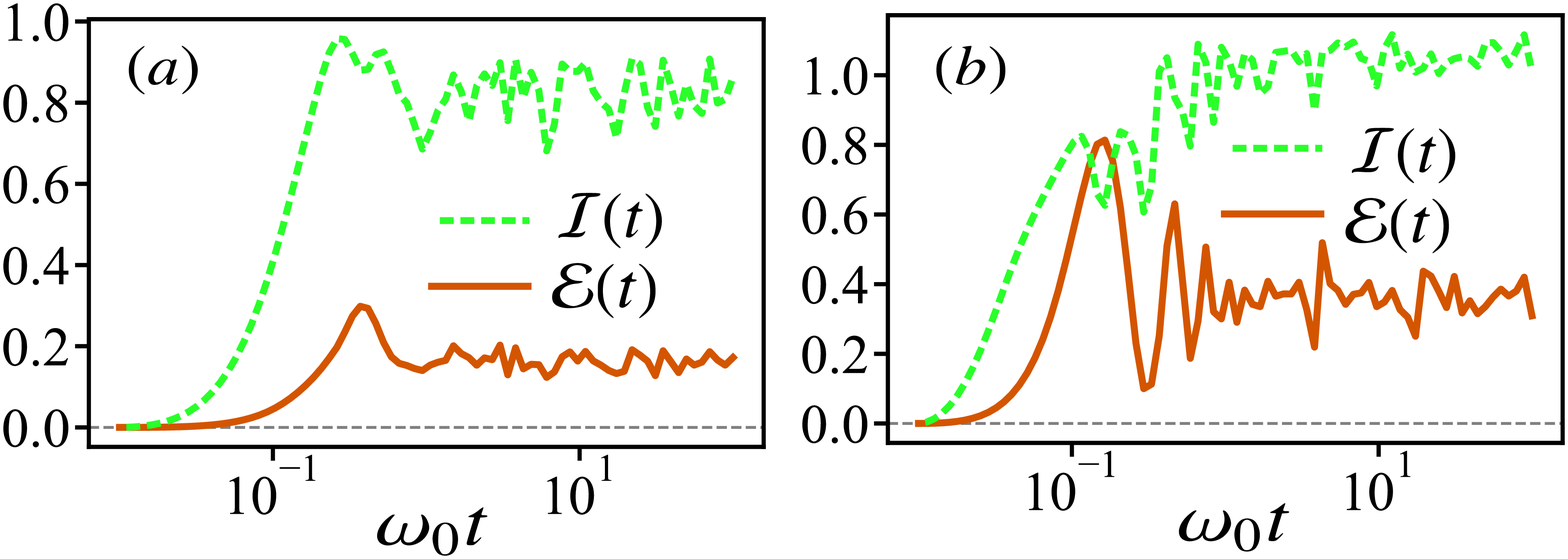}
	\caption{ Ergotropy of QB and battery-charger mutual information for (a) $\ket{\psi}$ and (b) $\ket{\phi^{+}}$ with $\alpha=2.5$.}
	\label{fig:mut-erg}
\end{figure}

\begin{figure}[h!]
	\centering
		\includegraphics[width=\linewidth]{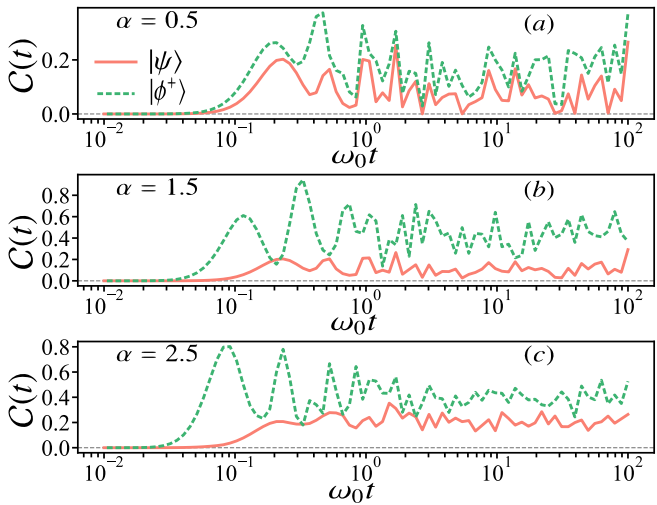}
	\caption{Quantum consonance of the QB with initially correlated and uncorrelated chargers (with the same parameter regimes as Fig.~\ref{fig:erg}).}
	\label{fig:conso}
\end{figure}

\begin{figure}[t!]
	\centering
		\includegraphics[width=\linewidth]{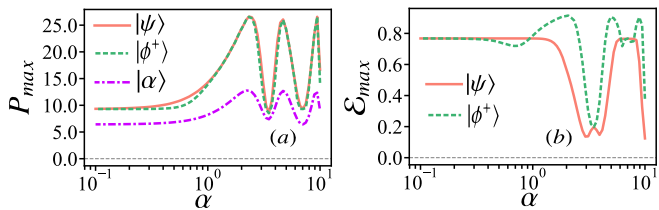}
	\caption{(a) Maximum amount of charging power versus $\alpha$ for single, double separable and double correlated chargers, (b) maximum of stored ergotropy of QB versus $\alpha$ for double separable and correlated chargers.}
	\label{fig:maxpowerg}
\end{figure}

We are also interested in the total amount of correlation between the QB and chargers i.e., mutual information $\mathcal I (t)=\mathcal I (\rho_{BC}(t))/N_{B}$ (Fig.~\ref{fig:ent-von}b). From Fig.~\ref{fig:ent-von} it is easy to show that $\mathcal I (t)=2 \mathcal S (t)$. This is the case for general pure states and implies the existence of both classical and quantum correlations between the QB and chargers. In fact, noting the Eq.~\ref{mutual}, it is well-known that for pure states quantum entanglement and quantum discord are the same ~\cite{al2011comparison} so $\mathcal S (t)$ can quantify the total amount of quantum correlations ($\mathcal S(\rho_{BC})=Q(\rho_{BC})$). Therefore, as the total state of system (QB+chargers) remains pure during the time evolution, $\mathcal I (t)=2 \mathcal S (t)$ means that the QB and chargers share both quantum and classical correlations, each one being the same and having the half part of total correlations (mutual information), meaning that $\mathcal S(\rho_{BC})=\chi(\rho_{BC})$ ~\cite{Henderson:01}. Moreover, all previous discussion for the Von Neumann entropy holds true for the mutual information as well. Comparing Figs.~\ref{fig:ent-von}a and ~\ref{fig:ent-von}b with ~\ref{fig:erg}c, one can highly confirm that the generation of correlation between QB and chargers positively contributes to the deposition of ergotropy in the QB. To a certain extent, whenever the QB and chargers are more correlated, larger amounts of ergotropy is stored in the QB. In Fig.~\ref{fig:mut-erg} we plot the ergotropy of QB along with  battery-charger mutual information to make their relevance more apparent.\\
The dynamics of generated quantum consonance in the quantum battery is shown in Fig.~\ref{fig:conso}. As stated before, the battery is initially prepared to be in its pure ground state $\rho_{B}(0)=\ket{g}^{\otimes 4}\bra{g}$ without any local and global coherence or correlations between qubits. Then, QC is generated in the QB due to the simultaneous interaction of the qubits with the chargers. Generally, both types of cavity's initial states are able to create QC among qubits of the QB. However, if one prepares the correlated state $\ket{\phi^{+}}$ as the chargers, larger amount of QC appears in the QB similar to the stored ergotropy. 

\section{Conclusions}
In this paper, we consider a 4-cell QB which can interact with one or two photonic cavities as chargers. We demonstrate that the simultaneous utilization of two independent chargers in coherent states considerably increases the charging power of the QB. This result is achieved while the single and double chargers scenarios contain the same average number of photons.\\
Moreover, It is shown that when the chargers are initially correlated namely being in a semi Bell state and by using a sufficient amount of fields intensity, more ergotropy and the ratio of useful extractable energy are stored in the QB in comparison to the case in which the chargers are in a product state without any correlations. We make this more evident by preparing the results for various types of semi Bell states. These results provide a good motivation for considering the role of different correlations between the chargers in the energy storage.\\
On the other hand, the QB with correlated chargers is less pure than that with uncorrelated ones, during the time evolution. However, the former results in more energy and ergotropy storage in the QB.\\
In addition, as the state of system (QB+chargers) undergoes a unitary evolution, it remains pure and we show that the total correlation between QB and chargers comprises both classical and quantum types with the same values. Finally, we see that the battery-charger correlations and non-local coherence of the battery positively interplay with the performance of the QB.
\appendix
\section{Maximum amount of ergotropy and power versus field's intensity}
\label{appendix:a}
It is already seen that (Fig.~\ref{fig:pow}) the charging power of QB benefits from double chargers (uncorrelated) for all values of $\alpha$. Now in this section we explore the dependence of maximum charging power ($P_{max}=\max_{t}[P(t)]$) and ergotropy ($\mathcal E _{max}=\max_{t}[\mathcal E (t)]$) on the field's intensity $\alpha$. Fig.~\ref{fig:maxpowerg}a shows $P_{max}$ versus $\alpha$ for three different initial states of chargers. The double charger cases appear with similar performances, no matter being correlated or not. This is why the result for charging power with correlated charger is not presented in the manuscript. What is more, a periodic dependence of maximum power on $\alpha$ is observed for all initial instances of chargers. While double chargers result in considerable advantage over the single charger, regarding the maximum power, the amount of such advantage oscillates with the degree of intensity. A similar discussion holds true for maximum ergotropy of QB over time evolution as shown by Fig.~\ref{fig:maxpowerg}b. Almost for all degrees of field's intensity, the initially correlated chargers are able to enhance the the maximum amount of deposited ergotropy in QB compared to the uncorrelated ones but the amount of this enhancement is $\alpha$-dependent.
\begin{figure}[t!]
	\centering
		\includegraphics[width=\linewidth]{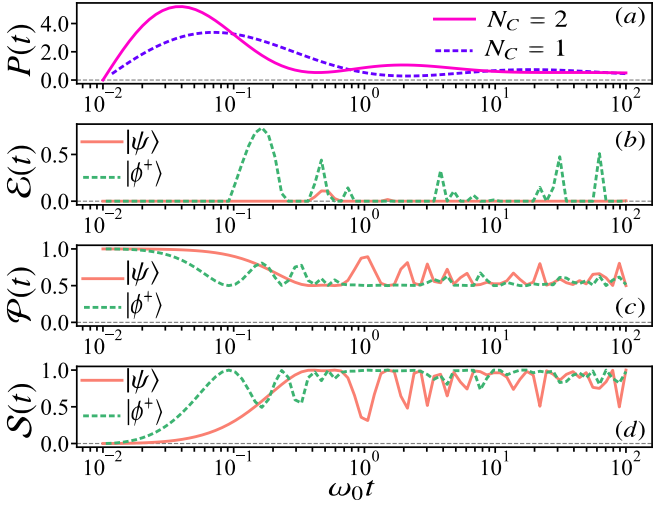}
	\caption{Dynamics of (a) charging power, (b) ergotropy, (c) purity and (d) Von Neumann entropy for a single qubit QB with chargers in states $\ket{\psi}$ and $\ket{\phi^{+}}$ for $\alpha=2.5$.}
	\label{fig:onequbit}
\end{figure}
\\
\begin{figure}[t!]
	\centering
		\includegraphics[width=\linewidth]{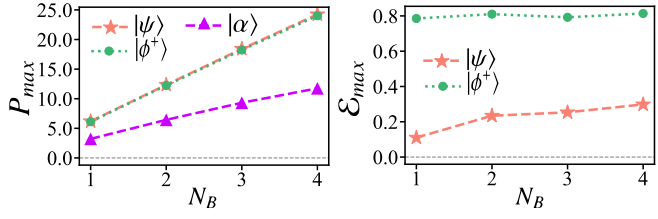}
	\caption{Maximum of (a) charging power and (b) ergotropy as a function of $N_{B}$ with $\alpha=2.5$.}
	\label{fig:scaling}
\end{figure}
\\
\begin{figure}[t!]
	\centering
		\includegraphics[width=\linewidth]{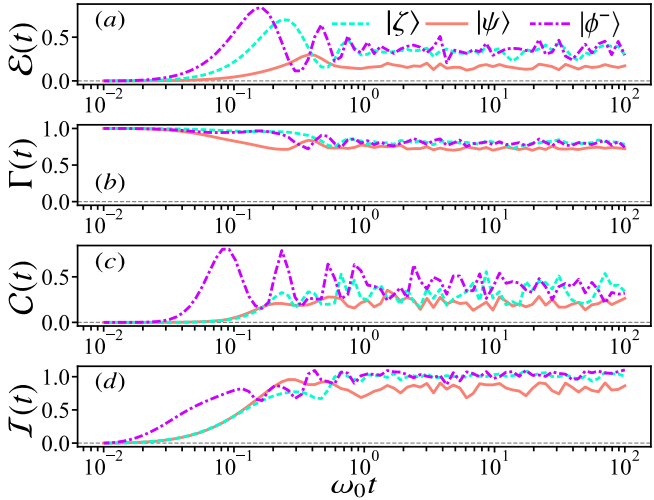}
	\caption{Dynamics of (a) ergotropy, (b) ratio of extractable energy, (c) quantum consonance of QB and (d) battery-charger mutual information with chargers in states $\ket{\psi}$, $\ket{\zeta}$ and $\ket{\phi^{-}}$ for $\alpha=2.5$.}
	\label{fig:others}
\end{figure}
\\
\begin{figure}[t!]
	\centering
		\includegraphics[width=\linewidth]{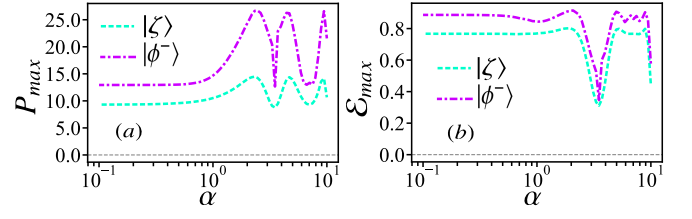}
	\caption{Maximum of (a) charging power and (b) ergotropy of QB as a function of $\alpha$ for correlated chargers $\ket{\zeta}$ and $\ket{\phi^{-}}$ with $\alpha=2.5$.}
	\label{fig:others2}
\end{figure}

\section{Scaling of maximum power and ergotropy with the size of QB}
\label{appendix:b}
It can be interesting to see the scaling trend of maximum power and ergotropy with the number of embedded cells (qubit). But first, let us check our results for a single-cell QB. For the sake of brevity, we only consider the results for $\alpha=2.5$. As seen from Fig.~\ref{fig:onequbit}, it can be seen that almost all previous results for the 4-cell QB are also valid here, one exception being the inability of the separable chargers in ergotropy storage. Next, we plot the maximum power and ergotropy as a function of QB's size ($N_{B}$) in Fig.~\ref{fig:scaling}. As shown by Fig.~\ref{fig:scaling}a, enlarging the QB provides faster charging process since the maximum power grows when $N_{B}$ is increased for all initial states of charger(s). The increasing rate of maximum charging power is the same for double correlated and uncorrelated chargers. Moreover, double chargers scheme offer a sharper growth of maximum power with $N_{B}$, compared to the single charger case. On the other hand, different scaling behavior is observed for maximum amount of ergotropy (Fig.~\ref{fig:scaling}b). Comparing the results for correlated and uncorrelated chargers, while the former demonstrate a far better performance for all $N_{B}$, for this case the maximum ergotropy is almost fixed over varying system's sizes. For the uncorrelated chargers, however, it shows a jump from $N_{B}=1$ to $N_{B}=2$ and then, slightly increases for other numbers of embedded qubits. We should note that in order to decide about the scaling trend with more certainty, one needs to go beyond such limited numbers of qubits however, we can at least rely on our results for these considered system sizes. 

\section{Other types of semi Bell states}
\label{appendix:c}
In this section, we provide complementary materials to prove the universality of our results. We consider other types of semi Bell states as the initial state of the photonic cavities namely, $\ket{\phi^{-}}=\frac{1}{\sqrt{N_{-}}}(\ket{\alpha}\ket{\alpha}-\ket{-\alpha}\ket{-\alpha})$ and $\ket{\zeta}=\frac{1}{\sqrt{\kappa}}(\ket{\alpha}\ket{0}+\ket{0}\ket{\alpha})$ with $N_{-}=2(1-e^{-4|\alpha|^{2}})$ and $\kappa=2(1+e^{-|\alpha|^{2}})$ ~\cite{mishra2016quantum}. As shown by Fig.~\ref{fig:others}, both $\ket{\phi^{-}}$ and $\ket{\zeta}$ store higher amounts of ergotropy and QC in the QB compared to $\ket{\psi}$, with $\ket{\phi^{-}}$ being the best choice. In the other words, when the chargers are initially correlated, one can see better performance of the QB regarding the stored ergotropy and generated quantum correlation. This is even the case if one considers the useful extractable energy $\Gamma$ Fig.~\ref{fig:others}b. Here, $\ket{\zeta}$ initially deposits a larger value of $\Gamma$, though in the steady limit $\ket{\phi^{-}}$ and $\ket{\zeta}$ have no considerable advantage over each other and both result in greater $\Gamma$ compared to uncorrelated initial chargers. For these two states also we plot the maximum of power and ergotropy as a function of $\alpha$ in Fig.~\ref{fig:others2}. A similar periodic dependence on $\alpha$ as for previous semi Bell state ($\ket{\phi^{+}}$) is apparent. Overall, $\ket{\phi^{-}}$ emerges with the best performance among the semi Bell states, regarding both the maximum of power and ergotropy.




\bibliography{mybib-URL.bib}

\end{document}